\documentclass[superscriptaddress,prl,twocolumn,aps]{revtex4-1}
\usepackage{amsmath,bm}
\usepackage{graphicx}
\usepackage{amssymb}
\usepackage{color}
\usepackage{lipsum}
\usepackage{xr}
\usepackage{soul}

\newcommand*\diff{\mathop{}\!\mathrm{d}}

\newcommand{\beq}{\begin{equation}}
\newcommand{\eeq}[1]{\label{#1}\end{equation}}
\newcommand{\brk}[1]{\left(#1\right)}
\newcommand{\Brk}[1]{\left[ #1 \right]}
\newcommand{\BRK}[1]{\left\{ #1 \right\}}

\newcommand{\matrixII}[4]{\left(\begin{array}{cc}#1&#2\\#3&#4\end{array}\right)}

\newcommand{\xhat}{\hat{\mathbf{x}}}
\newcommand{\Hvec}{\mathbf{H}}
\newcommand{\figref}[1]{Fig.~\ref{#1}}

\newcommand{\Kim}{{K}_\text{Im}}

\newcommand{\A}{\mathcal{A}}

\newcommand{\xvec}{\mathbf{x}}

\newcommand{\Qvec}{\mathbf{Q}}

\begin{document}
\title{Porous mechanical metamaterials as interacting elastic charges}

\author{Gabriele Librandi}
\thanks{G.L. and  M.M. contributed equally to this work}
\affiliation{Harvard John A. Paulson School of Engineering and Applied Sciences, Harvard University,
Cambridge, Massachusetts 02138, USA}
\author{Michael Moshe}
\thanks{G.L. and  M.M. contributed equally to this work}
\affiliation{Physics Department, Harvard University, Cambridge Massachusetts , USA}
\affiliation{Department of physics, Soft matter program, Syracuse University, New-York , USA}
\author{Yoav Lahini}
\affiliation{Harvard John A. Paulson School of Engineering and Applied Sciences, Harvard University,
Cambridge, Massachusetts 02138, USA}
\affiliation{Raymond and Beverly Sackler School of Physics and Astronomy,
Tel Aviv University, Ramat Aviv, Tel Aviv 69978, Israel}
\author{Katia Bertoldi}
\thanks{Correspondence to bertoldi@seas.harvard.edu}
\affiliation{Harvard John A. Paulson School of Engineering and Applied Sciences, Harvard University,
Cambridge, Massachusetts 02138, USA}
\affiliation{Kavli Institute, Harvard University, Cambridge, MA 02138}

\begin{abstract}
We present an analytical framework to describe the complex nonlinear response of two-dimensional porous mechanical metamaterials. We adopt a geometric approach to elasticity in which pores are represented by elastic charges, and show that this method captures with high level of accuracy both the shape deformation of individual pores as well as collective deformation patterns resulting from interactions between neighboring pores. Specifically, we show that quadrupoles and hexadecapoles - the two lowest order multipoles available in elasticity - are capable of matching the experimentally observed evolution in shape and relative orientation of the holes in a variety of periodic porous mechanical metamaterials. Our work demonstrates the ability of elastic charges to capture the physics of highly deformable porous solids and paves the way to the development of new theoretical frameworks for the description and rational design of porous mechanical metamaterials.
  
\end{abstract}

\maketitle

 Mechanical metamaterials  attempt to achieve properties that do not occur in natural systems through aggregates of building blocks that deform cooperatively in response to mechanical forces.  While the initial efforts in the field have focused on 
 mechanical properties in the linear regime \cite{kadic2013metamaterials,Christensen2015}, more recently nonlinearities and instability have emerged as promising tools to achieve new functionalities \cite{Martin2017}.  In particular, it has been shown that the dramatic changes in pattern induced by the applied deformation in elastic blocks perforated with periodic arrays of holes provide a good platform for the design of  mechanical metamaterials with switchable \cite{Bertoldi2010,Overvelde1,Bertoldi2008,Johnson2009,Jongmin_2013,Shu2016_EML} and programmable \cite{Florijn2014} response. A prominent example is that of a square array of circular holes in an elastic matrix \cite{MullinPRL,BertoldiJMPS,YangNanoL}. In such system buckling triggers the formation of a pattern of mutually orthogonal elongated pores, which has been exploited to design materials with negative Poisson's ratio \cite{Bertoldi2010} and tunable acoustic properties \cite{Bertoldi2008,BertoldiPRB2,Wang2013}. However,  the rational design of such porous structures remains an issue, since  the prediction of their non-linear response is highly non-trivial. 
 
In an attempt to capture with high fidelity both the macroscopic response and the local changes in geometry induced by the applied load in highly deformable porous metamaterials, non-linear Finite Element simulations have been successfully used \cite{MullinPRL,BertoldiJMPS}.  However, these analyses are computationally expensive and their cost become prohibitive as the size of the systems increase. To overcome this issue, a simple  model based on the interaction of dislocation dipoles has been proposed \cite{YangNanoL,Matsumoto2009}. While such model correctly captures the orientational order of the pattern forming upon loading, it does not  predict the deformed shape of the pores and the transition in the response induced by the instability.

In this Letter, we adopt a geometrical formulation of elasticity \cite{Koiter1966,Efrati2009} and propose a new analytical tool based on the formalism of elastic charges to describe the behavior of highly deformable porous mechanical metamaterials.
Elastic charges are singular sources of stress that have been shown to be capable of describing a variety of phenomena, including defects~\cite{moshe2015PNAS}, plastic events~\cite{moshe2016PH} and mechanical interactions between contractile cells~\cite{moshe2015PRE}, and of capturing geometric non linearities \cite{moshe2014ISF}.
Here, we demonstrate that they are also a promising tool to describe the mechanics of porous mechanical metamaterials, since they capture with high accuracy the experimentally observed deformation and relative orientation of the holes for a variety of pore shapes and arrangements. As such, we believe  that this formalism opens avenues for the development of new analytical tools that can facilitate the design of systems capable of achieving a targeted response.

Our approach is inspired by the well known method of image charges in electrostatics \cite{jackson1975}.
According to this method,  
conductive shells can be replaced by some charge distribution - called image charges - that preserves the appropriate boundary conditions and also obeys Poisson's equation. 
For example, the electric field around a spherical conductive shell subjected to a remote uniform external field can be found by replacing the sphere with a single dipole located at its center.
Analogously, instead of solving the full elastic boundary value problem, we show that the mechanics of highly deformable porous solids can be captured by placing an image elastic charge $K_\text{Im}$ inside each hole.
Explicitly, such image charges act as sources for the stress function $\psi$, which in the linear approximation satisfies
\begin{equation}
\frac{1}{Y} \Delta \Delta \psi = K_\text{Im},
\label{eq:BiLap}
\end{equation}
where $Y$ is the Young's modulus of the bulk material and ${\Delta}$ is the Laplace operator in the reference Lagrangian frame.
The stress $\sigma$ and strain $\varepsilon$ can  then be derived as
\begin{equation}
\begin{split}
\sigma^{\alpha\beta} &=  E^{\alpha\mu} E^{\beta\nu} \nabla_{\mu\nu}\psi, \\
\varepsilon_{\alpha\beta} &= \A_{\alpha\beta\gamma\delta} \sigma^{\gamma\delta},
\end{split}
\label{eq:StressStrain}
\end{equation}
where $\A$ is the elastic tensor,  ${E}$ is the anti-symmetric tensor and we have adopted the Einstein summation convention with indices running through $\BRK{1,\,2}$. 
Since monopoles and dipoles charges are topologically prohibited \cite{moshe2013ARMA},  the image  charge induced by an external field is at least of quadrupolar order. It follows that the leading terms of the image charges that are compatible with the symmetry of uniaxial loading are 
\begin{equation}
\begin{split}
\Kim = &Q^{\alpha\beta}\nabla_{\alpha\beta}\delta\brk{\xvec} + H^{\alpha\beta\gamma\delta}\nabla_{\alpha\beta\gamma\delta}\delta\brk{\xvec} + \dots
\end{split}
\label{eq:ImageCharge}
\end{equation}
where $\xvec$ is the position in Lagrangian frame and $\delta\brk{\xvec}$ is the Dirac delta function.
The quadrupole $\Qvec$ has three degrees of freedom and can be written as
\begin{equation}
\Qvec = p \matrixII{1}{0}{0}{1} + q \matrixII{\cos 2\theta_q}{\sin 2\theta_q}{\sin 2\theta_q}{-\cos 2\theta_q}
\end{equation}
where $p$ is an isotropic charge, $q$ is the magnitude of the pure quadrupole and $\theta_q$ is its orientation. Differently, the hexadecapole $\Hvec$ has four degrees of freedom, but here, 
at the lowest level of approximation, we only consider the two with the same symmetry of the quadrupole, so that the only non-zero components are
\begin{equation}
\begin{split}
&H^{1111}=H^{2211}=-H^{1122}= - H^{2222}= h \cos  2\theta_h,\\
&H^{1112}=H^{1121}=H^{2212}=  H^{2221}= h \sin 2\theta_h,
\end{split}
\end{equation}
with $h$ and $\theta_h$ denoting the magnitude of the charge and its orientation, respectively. 
 
\begin{figure}
\includegraphics[width=\columnwidth]{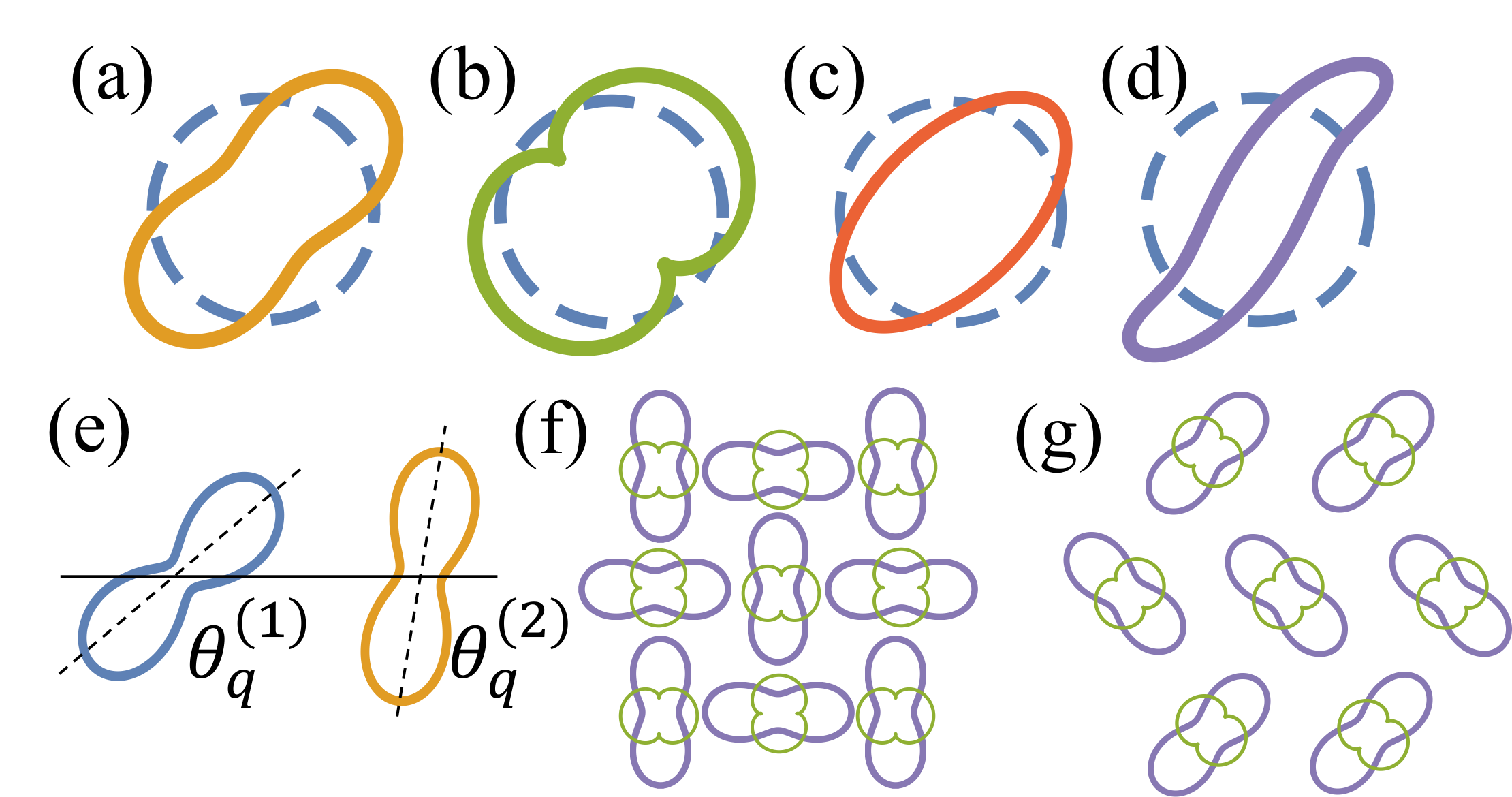}
\caption{Deformation of an isolated circular hole induced by  elastic charges characterized by various combinations of $p$, $q$, $h$, $\theta_q$ and $\theta_p$: (a) $q>0$, $p=h=0$ and $\theta_q =\pi/4$, (b) $h>0$, $p=q=0$ and $\theta_h =\pi/4$, (c) $h>0,\,q>0$, $p=0$ and $\theta_q=\theta_h =\pi/4$, and (d) $h>0,\,q>0$, $p=0$, $ \theta_q\gtrsim \pi/4$ and $\theta_h \lesssim \pi/4$. In all configurations the blue dashed line indicate the undeformed hole. (e) The interaction between two quadrupoles is minimized for $\theta_q^{(1)} + \theta_q^{(2)} = \pi/2$. (f) Energy minimizing configuration for a square array of quadrupoles (purple lines) and hexadecapoles (green lines) (g) Energy minimizing configuration for a triangular array of quadrupoles (purple lines) and hexadecapoles (green lines).}
\label{fig:Figure1}
\end{figure}
 
Next, we derive  the displacement field induced by the image charge associated to an isolated hole. To this end,  we  start by noting that  for a singular charge $\Kim = \delta\brk{\xvec}$ the stress function is  $\psi_{FS}\brk{\xvec} = \brk{Y/8\pi} |\xvec|^2 \brk{\ln |\xvec| - 1}$ . It follows that  $\psi$ for $\Kim$ as defined in Eq. (\ref{eq:ImageCharge})  can be calculated by simply superimposing  derivatives of $\psi_{FS}$ , yielding
\begin{equation}
\psi(\xvec) = \frac{Y}{8 \pi}\brk{ p \ln |\xvec| + 2  \, \xhat_{\alpha} \tilde{\Qvec}^{\alpha\beta} \xhat_{\beta} - 8 \frac{\xhat_{\alpha}\xhat_{\beta} \Hvec^{\alpha\beta\gamma\delta} \xhat_{\gamma} \xhat_{\delta}}{|\xvec|^2} }, 
\label{eq:psi}
\end{equation}
where $\tilde{\Qvec}$ is a pure quadrupole with no isotropic part, and $\xhat = \xvec/|\xvec|$. 
By substituting Eq. \eqref{eq:psi} in Eqs. \eqref{eq:StressStrain} and integrating for the displacement field we find

\begin{equation}
\begin{split}
u_r &= -\frac{p \brk{1+\nu}}{2\pi r} + \frac{q}{ \pi r}\cos \tilde{\theta}_q - \frac{2 h \brk{1+\nu}}{ \pi r^3}\cos \tilde{\theta}_h,\\
u_\theta &= \frac{q (\nu-1)}{ 2\pi r}\sin \tilde{\theta}_q - \frac{2h (1+\nu)}{ \pi r^3}\sin \tilde{\theta}_h,
\end{split}
\label{eq:QuadDisp}
\end{equation}
where $\tilde{\theta}_\alpha=2\theta-2\theta_\alpha$ (with $\alpha=h,q$), $(r,\theta)$ are polar coordinates in the Lagrangian frame, and $\nu$ is the Poisson's ratio of the bulk material.

In \figref{fig:Figure1}(a)-(d)  we show the calculated displacement field induced on an isolated circular hole by various combinations of $p$, $q$, $h$, $\theta_q$ and  $\theta_h$.  The similarity between these shapes and the shapes of deformed holes typically seen in highly deformable porous metamaterials \cite{BertoldiJMPS,Jongmin_2013} is remarkable and suggests us the capability of elastic charges to describe the mechanics of such systems.

From a theoretical perspective, to obtain the response for a metamaterial comprising an array of holes, one can use the solution for a single hole given by Eq. \eqref{eq:psi},  calculate the non linear corrections according to \cite{moshe2014ISF}, and compute the non linear energy of the system in terms of the interacting elastic charges located at each hole. While minimization of such energy with respect to the charges is expected to capture the response of  the metamaterial before and after the instability,  this procedure is very challenging. 
Here, we take an alternative path and perform two different tests to validate the capability of elastic charges to describe the behavior of porous mechanical metamaterials. First, we test the ability of Eq. \eqref{eq:QuadDisp}, which was calculated for an isolated hole, to describe the boundary deformation  of many holes embedded in a  structure. Second, we verify whether the orientation of neighboring holes is consistent with the interactions between the corresponding charges. 

At this point, it is important to point out that our approach (based on  interactions between charges in the linear approximation to describe the observed deformed patterns in metamaterials) can be explained in the spirit of Landau theory of phase transitions, where nonlinear interactions are encoded in renormalized coupling constants. In fact, the linear elastic energy written in terms of elastic charges accounts for two contributions: interactions between charges and interaction of charges with the external loads. While in the initial linear elastic regime (i.e. before buckling)  this energy has a single minimizer with all quadrupoles aligned with the external load, to capture the instability and the subsequent pattern transformation an effective linear description can be taken with renormalized coupling constants,  which are functions of the external loads and reflect higher order nonlinear interactions. 
Within this interpretation, the experimentally observed instability implies for a transition from a regime where external load dominates to a regime where interactions between different charges dominate. As such, post buckling configurations are expected to be described by  the charges orientations that minimize their linear interaction energy.

More specifically, if we consider  two image charges   $\Kim^{(1)}$ and $\Kim^{(2)}$, their interaction energy in the linear limit can be calculated as  
\begin{equation}
\label{eq:Uinteraction}
U = \int \psi^{(1)} \Kim^{(2)} \diff S = \int \psi^{(2)} \Kim^{(1)} \diff S,
\end{equation}
where $\psi^{(1)}$ and $\psi^{(2)}$ are the stress functions induced by the two charges. Substitution of Eq. \eqref{eq:ImageCharge} and Eq. \eqref{eq:psi} into Eq. \eqref{eq:Uinteraction} yields
\begin{equation}
U=U_{QQ}+U_{HH}+U_{QH},
\end{equation}
where
\begin{equation}
\begin{split}
U_{QQ}  &= \frac{Q^{(1)} Q^{(2)}}{\pi d^2} \cos \brk{2 \theta_q^{(1)} +2 \theta_q^{(2)}}, \\
U_{HH}  &= 0,\\
U_{QH}  &=  -\frac{12 Q^{(1)} H^{(2)}}{\pi d^4} \cos \brk{2 \theta_q^{(1)} +2 \theta_h^{(2)}}, 
\end{split}
\label{eq:Int}
\end{equation}
and $d$ denotes the distance between the two charges. Focusing on the two charges $\Kim^{(1)}$ and $\Kim^{(2)}$, Eq. \eqref{eq:Int} clearly indicate that: ($i$) two neighboring quadrupoles of fixed  magnitude and varying orientations, minimize their interaction energy $U_{QQ}$ if $\theta_q^{(1)}+\theta_q^{(2)} = \pi/2$ (see \figref{fig:Figure1}(e)); ($ii$) two neighboring hexadecapoles do not interact with each other; ($iii$) a quadrupole minimizes the interaction energy with the neighboring hexadecapole if $\theta_q^{(1)} + \theta_h^{(2)} = 0$. Similar arguments can be easily extended to predict the relative orientations of deformed holes embedded on  2D arrays. For example, assuming a square array of holes and taking into account interactions of each hole with its $8$ nearest neighbors,  we find that $U_{QQ}$ and $U_{QH}$ are minimized when the quadrupoles and hexadecapoles form a checkerboard pattern with $\theta_q = \Brk{0,\pi/2}$  and $\theta_h = \theta_q + \pi/2$, as shown in \figref{fig:Figure1}(f). 
Differently, when we consider holes on a triangular lattice and take into account interactions with the $6$ nearest neighbors, we obtain a zig-zag pattern with neighboring lines of charges alternatively oriented at $\theta_q =-\theta_h \approx \pm \pi/4$, as illustrated in \figref{fig:Figure1}(g).

Next, to verify the capability of elastic charges to capture the orientations and deformations occurring in porous mechanical metamaterials, we fabricate and experimentally test  three different samples. All our specimens are fabricated by lasercutting a periodic array of holes out of a polyurethane foam sheet (PORON 4701-40 Soft from Rogers Corporation with Poisson's ratio $\nu$=0.31) and compressed uniaxially under quasi-static conditions using a uniaxial loading frame with a 0.044 kN load cell 
in a displacement-controlled manner. The evolution of the holes is monitored  using a digital camera (Nikon D90 SLR) focused on the central holes both to reduce  boundary effects and to achieve  higher resolution.  The captured videos are then analyzed by digital image processing
(MATLAB) to extract the boundaries of the holes, from which their displacement field is reconstructed  (see Supplemental Material (SM) for details \cite{SI}). Finally, for each hole and at each level of considered applied deformation, we determine the combination of $p$, $q$, $h$, $\theta_q$ and $\theta_h$ in Eq. \eqref{eq:QuadDisp} that results in the displacement field  that best approximate the experimental one.

\begin{figure}
\includegraphics[width=\columnwidth]{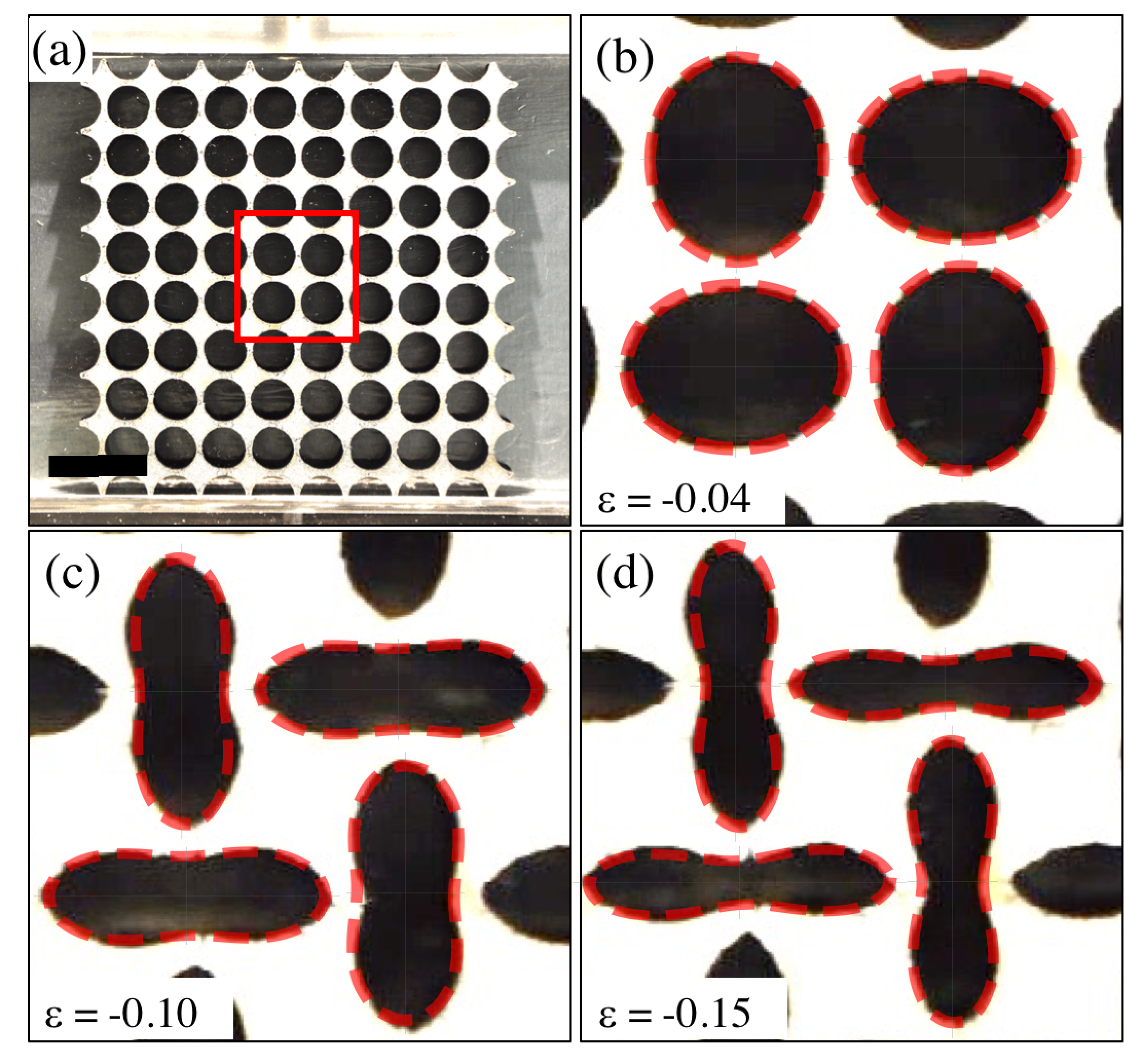}
\caption{(a)  Sample in the initial undeformed configuration. The red box indicate the four central holes we focused on in our analysis. Scale bar: 20 mm. (b)-(d) Experimental images of the four central holes  at different levels of macroscopic
strain: (b) -0.04, (c) -0.10 and (d) -0.15. The red dashed line indicate the best-fitting hole shapes generated by the elastic charges. }
\label{fig:Figure2}
\end{figure}

We start by  considering  a mechanical metamaterial comprising a 8$\times$8 array of circular holes with initial radius $r_0=4.18$ mm and center-to-center spacing $d=10$ mm, resulting in an initial porosity $\Psi=\pi r_0^2/d^2=0.55$.  In Figs. \ref{fig:Figure2}(b)-(d) we present experimental images of the sample (focusing on the four central holes included in the red box in Fig. \ref{fig:Figure2}(a)) at different levels of applied macroscopic strain.   As previously observed \cite{MullinPRL,YangNanoL,BertoldiJMPS},  when such sample is uniaxially compressed an elastic instability results in a sudden transformation of the circular pores to a periodic pattern of alternating, mutually orthogonal, elongated holes (see Figs. \ref{fig:Figure2}(b)-(d) and Movie S1 \cite{SI}). In the snapshots shown in Fig. \ref{fig:Figure2} we also indicate with red dashed lines the best-fitting hole boundaries  generated by the elastic charges using Eq. \eqref{eq:QuadDisp}. Notably, we find that the elastic charges are capable of matching extremely well the evolution in shape experienced by the pores and capture also the small details of the deformed pores, even in the presence of neighboring holes. While close to the buckling point the holes have an elliptical shape, further away from the instability they take a two-lobed shape, which can be still nicely described by Eq. \eqref{eq:QuadDisp}. 

\begin{figure}
\includegraphics[width=\columnwidth]{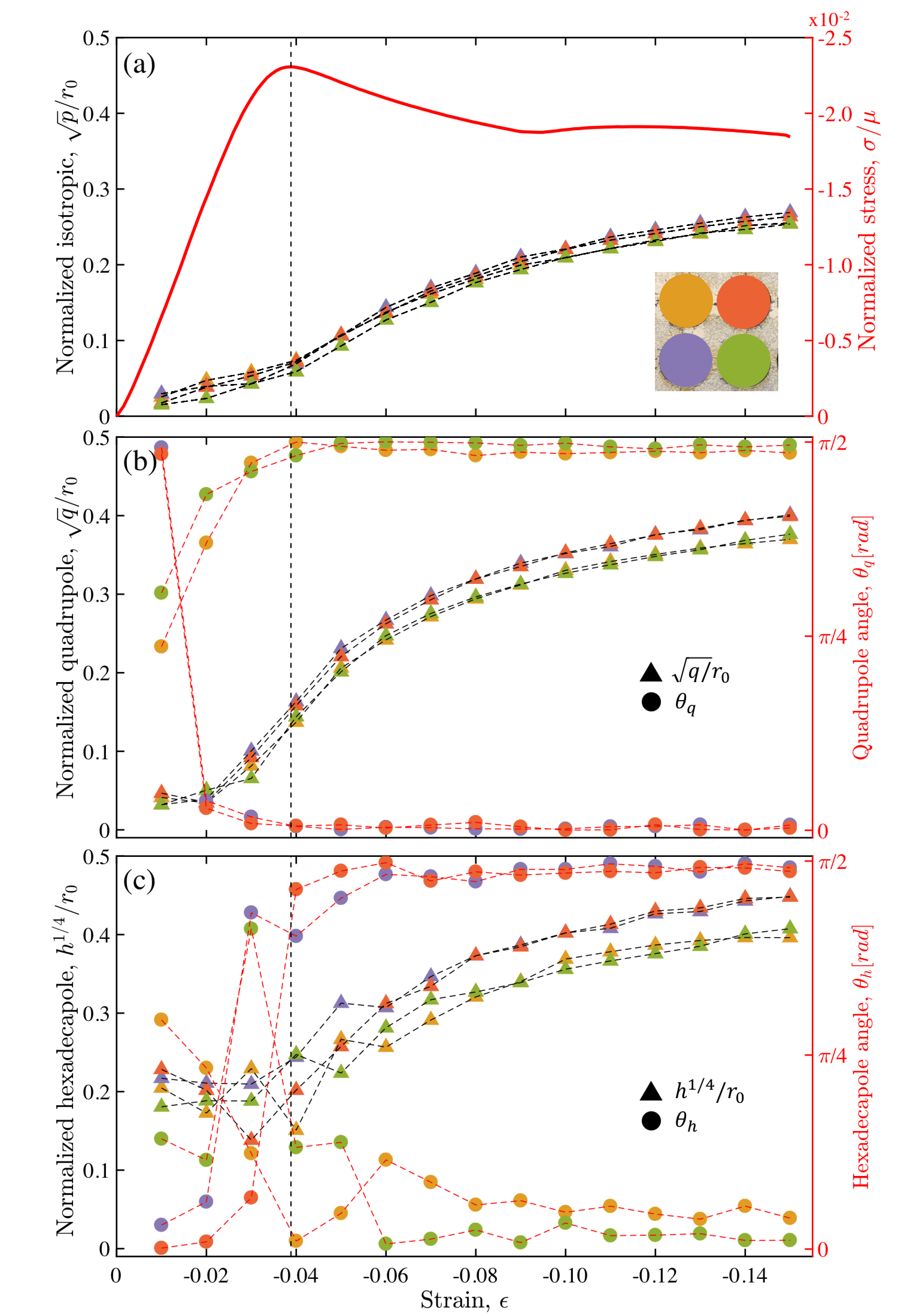}
\caption{ (a) Evolution of the normalized isotropic charges as a function of the applied strain for the four central holes (triangular markers - left axis); stress-strain curve (red line - right axis). (b) Evolution of the normalized magnitude (triangular markers - left axis) and orientation (circular markers - right axis) of the quadrupole charges as a function of the applied strain for the four central holes. (c) Evolution of the normalized magnitude (triangular markers - left axis) and orientation (circular markers - right axis) of the hexadecapole  charges as a function of the applied strain for the four central holes. }
\label{fig:Figure3}
\end{figure}

Next, we turn our attention to the  evolution of the charge magnitudes  and  orientations as a function of the applied strain. In \figref{fig:Figure3} we present the evolution of the normalized charge  magnitude   for the four central holes. In the same plots we also show the evolution of the angles $\theta_q$ and $\theta_h$ (\figref{fig:Figure3} (b) and (c) - right axes). Note that, since the dimensions of  charges are $[p]=[q] = L^2$ and  $[h]=L^4$ ($L$ standing for length), we define their normalized values as  $\sqrt{p}/r_0$, $\sqrt{q}/r_0$, and $ h^{1/4}/r_0$.  Our results indicate that for the considered highly porous structure all charges are equally important (i.e. $\sqrt{p} \sim \sqrt{q} \sim h^{1/4}$). We also find that  all charge magnitudes and orientations undergo a transition  at $\epsilon_c=-0.0388$, as highlighted by the vertical black dashed line in the plots. Note that at $\epsilon_c$  a plateau also emerges in the stress-strain curve (Fig. \ref{fig:Figure3}(a) - red line), indicating that  this is the critical strain at which  buckling  occurs. 
At the onset of buckling  not only $p$, $q$ and $h$  increase, but also  $\theta_p$ and $\theta_q$ suddenly approach either $0$ or $\pi/2$, leading to an alternate pattern with quadrupoles and hexadecapoles perpendicular to each other at each hole. 
This is fully consistent with the theoretical prediction for the relative orientations of neighboring charges illustrated in Fig.\ref{fig:Figure1}(f). Finally, we note that the magnitudes of the charges plotted in Fig. \ref{fig:Figure3} are slightly larger for the   holes aligned horizontally  than  for the vertical ones, implying that the interaction between the charges and the external load also plays a role. 

Having demonstrated that elastic charges correctly describe the mechanics of a metamaterial comprising a square array of circular holes, we now test their ability to capture the deformation of systems with different hole arrangement and shape.  To this end, we test two additional porous mechanical metamaterials. The first one still comprises a periodic array of circular holes, but arranged on a triangular lattice. As shown in  Figs.  \ref{fig:Figure4}(a)-(d) the elastic charges are able to capture with high fidelity the  zig-zag pattern induced by the instability as well as the non-trivial S-shape of the pores at high deformation. Note that this last feature cannot be captured without taking into account hexadecapole contributions (see Fig. S3 \cite{SI}).   As for the orientation of the charges, we find that for the holes of one row $\theta_q\sim-\theta_h\sim\pi/4$, while for those of the other one $\theta_q\sim-\theta_h\sim-\pi/4$ (see Fig. S4 \cite{SI}),  exactly as predicted by our analysis (see Fig. \ref{fig:Figure1}(g)). Moreover, as already observed for the case of square arrays of holes, also for this system the interaction with the external load plays an important role. While in the former it only slightly affects the magnitude of the charges, here  compression applied in perpendicular direction results in a completely different charge distribution (see  Fig. S5 \cite{SI}).

Finally, in  Fig. \ref{fig:Figure4}(e)-(h) we consider a structure   comprising a 8$\times$8 rectangular array of ellipses with aspect ratio $a/b=1.898$. In this system the applied uniaxial compression trigger an instability which induces the formation of a pattern of alternating elongated and almost circular holes \cite{BertoldiJMPS} (see Figs. \ref{fig:Figure4}(g)-(h)). Again,  we find that this deformed pattern can be nicely captured by elastic charges, as indicated in Figs. \ref{fig:Figure4}(f)-(h) by the red dashed lines,  which represent the best-fitting hole boundaries generated using Eq. \eqref{eq:QuadDisp}. {Furthermore, as for the case of the metamaterial with a square array of circular holes, we find that after buckling is triggered (i.e. at $\epsilon_c =-0.0519 $) $\theta_h=\theta_q\pm\pi/2$ (see Fig. S6 \cite{SI}). 
However, differently from the  two metamaterials considered above {for which all holes take similar shapes, for the structure with elliptical holes two families of holes emerge at $\epsilon>\epsilon_c$ and this is reflected in the values of the elastic charges (Fig. S6 \cite{SI}).

In summary, in this work we have implemented the framework of elastic charges to study the response of highly deformable porous mechanical metamaterials. Our results indicate that this formalism is capable of capturing with high level of accuracy the evolution of the boundary deformation and relative orientation  observed in elastic solids perforated with square and triangular arrays of circular holes as well as rectangular arrays of elliptical holes. 
An advantage of the proposed approach is the intuitive picture that it provides: thinking of deformed holes in a porous material as interacting elastic charges may be used to predict and design patterns of holes capable of targeted deformations without a direct calculation.
The main focus of this investigation has been to determine the capability of elastic charges to capture the physics of porous perforated solids. While the relative orientations between adjacent deformed holes have been obtained analytically, charges magnitudes have been found using a fitting procedure. 
The success of this study advocates a (future) deeper theoretical investigation of perforated solids based on minimization of the nonlinear energy of the system calculated in terms of the interacting elastic charges located at each hole. 
Finally, it is important to emphasize that the generalization of the elastic charge formalism to three-dimensional settings is not yet established. The present work shows that such a generalization may be highly valuable and open avenues for the design of 3D mechanical metamaterials with targeted response.

\begin{figure}
\includegraphics[width=\columnwidth]{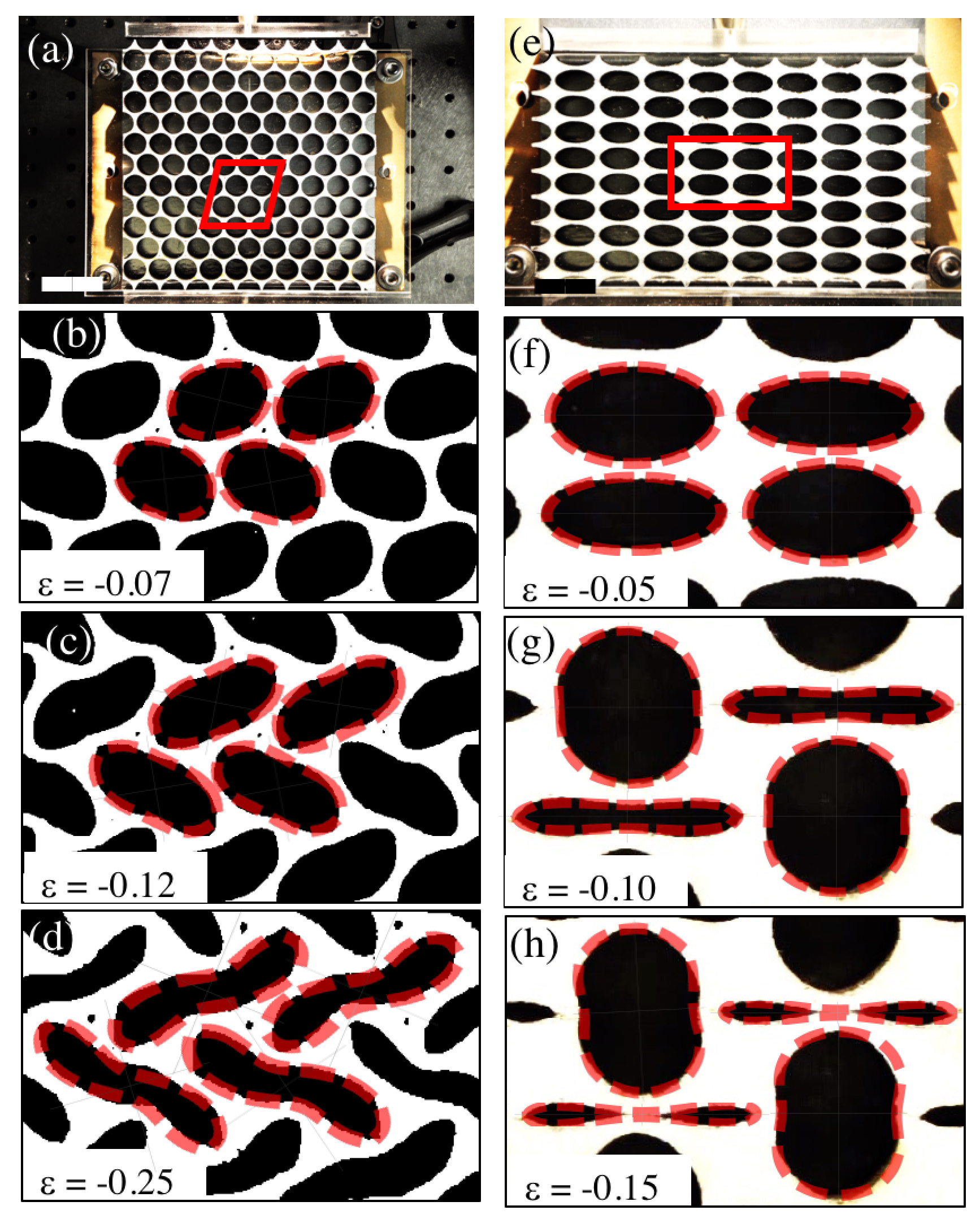}
\caption{(a) and (e) samples in the initial undeformed configuration. The red boxes indicate the holes we focused on in our analysis. Scale bar for the triangular lattice sample: 30 mm. Scale bar for the rectangular lattice sample: 20 mm. (b)-(d) and (f)-(h) experimental images of the four holes at different levels of macroscopic strains (see the insets). The red dashed lines indicate the best-fitting hole shapes generated by the elastic charges}
\label{fig:Figure4}
\end{figure}

\begin{acknowledgments}
KB acknowledges support from by the Materials Research Science and Engineering Center under NSF Award number DMR-1420570. MM acknowledges support from the NSF through the DMREF
grants DMR-1435794, DMR-1435999, and the USIEF Fulbright program. MM thank Eran Sharon, Raz Kupferman and Yohai Bar-Sinai for fruitful discussions.
\end{acknowledgments}

\bibliography{References}

\end{document}